\documentclass[fleqn,usenatbib]{mnras}
\usepackage{amsmath}
\usepackage{appendix}
\usepackage{graphicx}
\usepackage{float}
\usepackage{amssymb}
\usepackage{color}
\usepackage{float}

\usepackage{newtxtext,newtxmath}

\usepackage[T1]{fontenc}

\DeclareRobustCommand{\VAN}[3]{#2}
\let\VANthebibliography\thebibliography
\def\thebibliography{\DeclareRobustCommand{\VAN}[3]{##3}\VANthebibliography}

\title[HIC of photosphere model in different regimes]{The Hardness-intensity Correlation of Photospheric Emission from a Structured Jet for Gamma-Ray Bursts}

\author[X. Y. Song et al.]{
Xin-Ying Song$^{1}$\thanks{E-mail: songxy@ihep.ac.cn},
Yan-Zhi Meng$^{2,3}$
\\
$^{1}$Key Laboratory of Particle Astrophysics, Institute of High Energy Physics, Chinese Academy of Sciences, Beijing 100049, China\\
$^{2}$School of Astronomy and Space Science, Nanjing University, Nanjing 210023, China\\
$^{3}$Key Laboratory of Modern Astronomy and Astrophysics (Nanjing University), Ministry of Education, China\\
}

\date{Accepted 2022 March 22. Received 2022 February 20; in original form 2021 December 26}

\pubyear{2021}

\begin{document}
\label{firstpage}
\pagerange{\pageref{firstpage}--\pageref{lastpage}}
\maketitle

\begin{abstract}
 For many gamma-ray bursts (GRBs), hardness-intensity correlation (HIC) can be described by a power-law function, $E_{\rm p}\propto F^{\kappa}$, where $E_{\rm p}$ is the peak energy of $\nu F_{\nu}$ spectrum, and $F$ is the instantaneous energy flux. In this paper, HIC of the non-dissipative photospheric emission from a structured jet is studied in different regimes. An intermediate photosphere, which contains both of unsaturated and saturated emissions is introduced, and  we find positive $\kappa<1/4$ in this case. The same conclusion could be generalized to the photospheric emission from a hybrid jet without magnetic dissipations, or that with sub-photospheric magnetic dissipations and fully thermalized. This may imply that the contribution
peaking at $\sim1/2$ in the distribution of observed $\kappa$ are mainly from
the prompt emission of GRBs with synchrotron origin. Besides, emissions of the intermediate photosphere could give a smaller low-energy photon index $\alpha$ than that in the unsaturated regime, and naturally reproduce anti-correlation in $\alpha-E_{\rm p}$ in a GRB pulse.
\end{abstract}




\section{Introduction} \label{sec:intro}

Photospheric emission, as a natural consequence of the fireball model, can offer  an interpretation of the low-energy photon index ($\alpha$) of greater than -2/3~\citep[so-called `line of death',][]{Preece_1998} of the gamma-ray burst (GRB) spectra. The Planck spectrum related to photospheric emission is too narrow, and can be described by an exponential cut-off power law (CPL, also called Comptonized model) with $\alpha=1$, which is not a typical value for $\alpha$. However, it could be widened in two ways. Firstly, dissipation below the photosphere can heat electrons above the equilibrium temperature. These electrons emit synchrotron emission and comptonize thermal photons, thereby modify the shape of the Planck spectrum~\citep{2005ApJ...635..476P, Pe_er_2006, 2005Dissipative}. The observational evidence for the subphotospheric heating has been provided by \cite{2011Observational}. Besides, internal shocks bellow the photosphere~\citep{2005Dissipative}, magnetic reconnection~\citep{1994A, 2004Spectra}, and hadronic collision shocks~\citep{2010Collisional, 2010Radiative} can also cause dissipation. Secondly, the modification of the Planck spectrum could be caused by geometrical broadening. Photospheric radius is found to be a function of the angle to the line of sight~\citep{1991The, 2008ApJ...682..463P, 2018ApJ...860...72M}, therefore, the observed spectrum is a superposition of a series of blackbody of different temperatures, arising from different angles to the line of sight. A multicolor Blackbody (mBB) model introduced and formulated by~\cite{2010ApJ...709L.172R} and \cite{Hou_2018}, is utilized to describe the spectrum which is broader than the Planck spectrum with a single temperature. BAND or CPL model could describe this kind of spectra as well in some cases as discussed in \cite{Hou_2018}. 

\cite{2008ApJ...682..463P} shows that photons make their last scatterings at a distribution of radii and angles. Based on this model, the theory of photospheric emission from relativistic jets with angle-dependent outflow properties is developed in ~\cite{2013A}, and a structured jet with angle-dependent baryon loading parameter profiles is considered. Average low-energy photon index ($\alpha=-1$) could be obtained and independent of viewing angle.
In \cite{2019The}, the observed evolution patterns of the $\nu F_{\nu}$ peak energy ($E_{\rm p}$), including hard-to-soft and intensity-tracking, are reproduced by the non-dissipative photosphere (NDP) model with a structured jet considered. This implies that, it may be reasonable to study the evolution of GRB spectra from the emission of NDP model from a structured jet.

The hardness-intensity (HI) study started from 1983~\citep{1983Correlation}. The relation between the hardness and the intensity, during the prompt phase of GRBs, have been well investigated. It shows that there is no ubiquitous trend of spectral evolution that can characterize all bursts. Most cases exhibit a hard-to-soft behavior over a pulse, with the hardness decreasing monotonically as the flux rises and falls~\citep{1996ApJ...459..393N}, while a few cases show soft-to-hard, soft-to-hard-to-soft or even more chaotic evolution. Various types of trends may exist in a single GRB~\citep{1993ApJ...413..281B, 1995ApJ...439..307F}, and in recent results from~\cite{Li_2021} which consists of 39 bursts, 117 pulses and 1228 spectra, the general trend is confirmed that pulses become softer over time, with $\alpha$ becoming smaller. \cite{2000On} comprises a sample of 82 long pulses selected from 66 long bursts observed by the Burst and Transient Source Experiment (BATSE) on the Compton Gamma-Ray Observatory. It is found that at least $57\%$ of these pulses have HICs that could be described by a power law. A power-law relation between the instantaneous energy flux, $F$ (erg cm$^{-2}$ s$^{-1}$), and  $E_{\rm p}$~(keV) which serves as a measurement of the hardness is shown as  
\begin{equation}\label{eq:FvsE}
 E_{\rm p}\propto F^{\kappa},
\end{equation}
where $\kappa$ is the index of the power-law function. The bolometric flux is more intrinsic and suggested by~\cite{2000On}. Practically, $F$ and $E_{\rm p}$ are always represented by those of the time-averaged spectrum in each time interval in the time-resolved analysis. From~\cite{1983Correlation}, $index=1/\kappa$ in $F\propto E_{\rm p}^{index}$ is found to be a typical value of 1.5~$\sim$~1.7. \cite{2000On} presents a value of $index$ varied from 1.4 to 3.4 with a wider spread, with a mean of 1.9 and a standard deviation of 0.7. \cite{Lu:2012pf} shows the measurement with $\kappa=0.55\pm0.22$ for both long and short GRBs. These measurements are consistent well with each other, and it implies that there exists a characteristic value of $\kappa\sim 0.5$ or $index\sim2$.

 In this paper, the emission of the NDP model from a structured jet is considered, furthermore, the jet adopted here is dominated by the thermal energy, and the photospheric emission originates from the relativistic outflow with a pure hot fireball component. A motivation is raised that we wonder if this model in different regimes could reproduce or cover the range of observed $\kappa$. With the same method, the cases of the hybrid relativistic outflow with or without magnetic dissipation could be discussed as well. Patterns of $\alpha-E_{\rm p}$ in one GRB pulse are also extracted and discussed.

This paper is organized as follows. In Section~\ref{sec:modelintro}, assumptions of the jet structure and the NDP model are introduced. In Section~\ref{sec:emissions}, different regimes are discussed in detail; HICs and $\alpha-E_{\rm p}$ are extracted. In Section~\ref{sec:discuss}, results are discussed; the conclusions are drawn and generalized to the case of the hybrid relativistic outflow. The conclusions are summarized in Section~\ref{sec:sum}.  

\section{A structured Jet and Photosphere Model}\label{sec:modelintro}

As shown in \cite{2013A}, \cite{2018ApJ...860...72M} and  \cite{2019The}, the jet is structured with an inner-constant and outer-decreasing angular baryon loading parameter profile with the form
\begin{equation}\label{eta_theta}
(\eta(\theta) -\eta _{\min })^{2}=\frac{(\eta_{0}-\eta _{\min })^{2}}{%
(\theta /\theta _{c})^{2p}+1},
\end{equation}%
where $\eta$ is the baryon loading parameter which is also the bulk Lorentz factor $\Gamma$ in the saturated acceleration regime, $\eta_0$ is the maximum $\eta$, and also denoted as $\Gamma_0$, $\theta$ is the angle measured from the jet axis, $\theta_{c}$ is the half-opening angle for the jet core, $p$ is the power-law index of the profile, and $\eta _{\min }=1.2$ is the minimum value of $\eta$. An angle-dependent luminosity~\citep{2001Gamma,10.1046/j.1365-8711.2002.05363.x, 2002Gamma,2003THE} could also be considered in this analysis, and has the form 
\begin{equation}\label{eq:lumi_theta}
L(\theta)=\frac{L_{0}}{((\theta /\theta _{\rm c,L})^{2q}+1)^{1/2}},
\end{equation}%
where $\theta _{\rm{c,L}}$ is the half-opening angle for the luminosity core, while $q$ describes how the luminosity decreases outside the core. 

 Photospheric radius $R_{\rm ph}$ is defined as the radius where the scattering optical depth for a photon moving toward the observer is equal to unity ($\tau=1$). The photons can be scattered at any position ($r$, $\Omega$) inside the outflow in principle, where $r$ is the distance from the explosion center and $\Omega$ ($\theta$, $\phi$) is the angular coordinates. \cite{2013A} introduces the flux of observed energy $E_{\rm obs}$ at the observer time $t$ in the case of impulsive injection, and deduced as 
\begin{equation} \label{FEob}
\begin{split}
 F_{E}^{\rm obs}(\theta _{\rm{v}}, E_{\rm obs}, t) &=\frac{1}{4\pi d_{\text{L}}^{2}}
\int\int (1+\beta )D^{2}\frac{d\dot{N}_{\gamma }}{d\Omega }\times \frac{R_{
\text{dcp}}}{r^{2}}\\
&\exp \left( -\frac{R_{\text{ph}}}{r}\right)  \times \left\{ E\frac{dP}{dE}\right\}\times \frac{\beta c}{u} d\Omega dr, \\
&r=\frac{\beta c t}{u}, E=E_{\rm obs}(1+z),
\end{split}
\end{equation}
where the velocity $\beta=\frac{v}{c}$ and the Doppler factor $D=[\Gamma (1-\beta\cos \theta _{\text{LOS}})]^{-1}$ both depend on the angle $\theta $ to the jet axis of symmetry, in which $\theta _{\text{LOS}}$ is the angle to the line of sight (LOS) of the observer. The viewing
angle $\theta _{\text{v}}$ is the angle of the jet axis of symmetry to the LOS. $d\dot{N}_{\gamma }/d\Omega =\dot{N}_{\gamma }/4\pi $ and $\dot{N}%
_{\gamma }=L/2.7k_{\text{B}}T_{0}$, where $L$ is the total outflow luminosity, $T_{0}=(L/4\pi r_{0}^{2}ac)^{1/4}$ is the base outflow temperature and $a$ is the radiation constant. The angle-dependent decoupling radius $R_{\rm dcp}$, as the radius from which the optical depth for a photon that propagates in the radial direction is equal to unity, is defined as
\begin{equation}
R_{\text{dcp}}=\frac{1}{(1+\beta )\beta \Gamma ^{2}}\frac{\sigma _{\text{T}}
}{m_{\text{p}}c}\frac{d\dot{M}}{d\Omega },  \label{Rdcp}
\end{equation}
where $d\dot{M}(\theta )/d\Omega =L/4\pi c^{2}\eta (\theta )$ is the angle-dependent mass outflow rate per solid angle. $R_{\rm ph} $ is defined as
\begin{equation}\label{Rph}
R_{\text{ph}}=\frac{\sigma _{\text{T}}}{m_{\text{p}}c\sin \theta _{\text{LOS}
}}\int\nolimits_{0}^{\theta _{\text{LOS}}}\frac{(1-\beta \cos \tilde{\theta}
_{_{\text{LOS}}})}{\beta }\frac{d\dot{M}}{d\Omega }\text{ }d\tilde{\theta}
_{_{\text{LOS}}},
\end{equation}
and especially, $R_{\rm ph}= R_{\rm dcp}$ if photons propagate along the LOS, and $R_{\rm dcp}\sim R_{\rm ph}$ in lower latitude as shown in Figure 4 in \cite{2013A}. $dP/dE$ describes the probability for a photon to have an observer frame energy between $E$ and $E+dE$ within volume element $dV$, and it is a comoving Planck distribution
with the comoving temperature $T^{\prime}(r,\Omega)$, and can be written as
\begin{equation}\label{dPdE}
\frac{dP}{dE}=\frac{1}{2.40(k_{\text{B}}T_{\text{ob}})^{3}}\frac{E^{2}}{\exp
(E/k_{\text{B}}T_{\text{ob}})-1},
\end{equation}%
where $k_{\rm B}$ is the Boltzmann constant, $T_{\text{ob}}(r,\Omega )=D(\Omega )\cdot $ $T^{\prime }(r,\Omega )$
is the observer frame temperature, and in the saturated acceleration regime (the saturation radius is smaller than the photospheric radius, $R_{\rm s}<R_{\rm ph}$), it is defined by
\begin{equation}
T^{\prime }(r,\Omega )=\left\{
\begin{array}{c}
\frac{T_{0}}{\Gamma (\Omega )},\text{ \ \ \ \ \ \ \ \ \ \ \ \ \ \ \ \ \ \ \ }%
r<R_{s}(\Omega )<R_{\text{ph}}(\Omega ), \\
\frac{T_{0}[r/R_{s}(\Omega )]^{-2/3}}{\Gamma (\Omega )},\text{ \ \ \ \ \ \ \
}R_{s}(\Omega )<r<R_{\text{ph}}(\Omega ), \\
\frac{T_{0}[R_{\text{ph}}(\Omega )/R_{s}(\Omega )]^{-2/3}}{\Gamma (\Omega )},%
\text{ }R_{s}(\Omega )<R_{\text{ph}}(\Omega )<r.%
\end{array}%
\right.
\end{equation}
In the unsaturated acceleration regime ($R_{\rm s}>R_{\rm ph}$), as discussed in \cite{2018ApJ...860...72M} and \cite{2019The}, which always exists in lower luminosity or larger $r_0$, $\Gamma$ can not reach the value of $\eta$.
$d\dot{N}_{\gamma }/d\Omega $ is still calculated by $\dot{N}%
_{\gamma }=L/2.7k_{\text{B}}T_{0}$. 
For the case of $R_{\rm s}\gg R_{\rm ph}$, 
The decoupling radius $R_{\text{dcp}} (\sim R_{\rm ph}$) are given by \cite{2018ApJ...860...72M} and \cite{2019The},
\begin{equation}\label{Rdcp2}
R_{\text{dcp}}=\left[\frac{\sigma _{\text{T}}}{6m_{\text{p}}c}%
\frac{d\dot{M}}{d\Omega }r_{0}^{2}\right] ^{1/3}\text{.}
\end{equation}%
If $R_{\rm s}\gtrsim R_{\rm ph}$,
\cite{Meszaros_2000} gives the common form of $R_{\rm ph}$ as  
\begin{equation}\label{Rdcp3}
R_{\text{ph}}=\left[\frac{\sigma _{\text{T}}}{m_{\text{p}}c}%
\frac{d\dot{M}}{d\Omega }r_{0}^{2}\times \frac{Y}{\Gamma_{r_0}^2}\right] ^{1/3}\text{,}
\end{equation}%
where $Y$ denotes the number of electrons per
baryon and $\Gamma_{r_0}\gtrsim1$, is the initial bulk Lorentz factor at $r_0$. Besides, the comoving temperature $T^{\prime}(r)$ is given by
\begin{equation}\label{T2}
T^{\prime }(r)=T_{0}/\Gamma,
\end{equation}
where $\Gamma$ is $R_{\rm dcp}/r_0$.

Continuous wind could be assumed to consist of many thin layers from an impulsive injection (spectra are shown in Figure~\ref{fig:Lph} (a) and (c) for unsaturated and saturated emissions) at its injection time $\hat{t}$, and the wind luminosity at $\hat{t}$ is denoted as $L_{\rm w}(\hat{t})$. The flux of observed energy $E_{\rm obs}$ at the observer time $t$ is given by 
\begin{equation} \label{eq:FEob2}
\begin{split}
 &F_{E_{\rm{obs}}}^{\rm obs}(\theta _{\rm{v}}, E_{\rm obs}, t)=\\
 &\begin{cases}
\int\nolimits_{0}^{t}F_{E}^{\text{obs}}(\theta _{\text{v}},t,\hat{t}, L_{\rm w}(\hat{t}))d%
\hat{t},\textbf{\ \ \  }t< t_{\rm D}, \\
\int\nolimits_{0}^{t_{\rm D}}F_{E}^{\text{obs}}(\theta _{\text{v}},t,\hat{t}, L_{\rm w}(\hat{t}))d%
\hat{t},\textbf{\ \ \  }t\geq t_{\rm D},\\
\end{cases}
\end{split}
\end{equation}
where $t_{\rm D}$ is the duration of emission of the central engine, and $r=\frac{\beta c (t-\hat{t})}{u}$ in $F_{E}^{\text{obs}}(\theta _{\text{v}},t,\hat{t}, L_{\rm w}(\hat{t}))$ in Equation~(\ref{FEob}). The observed GRB has a duration, thus, it is reasonable that the central engine produces a continuous wind. For the case of the constant wind luminosity, after the central engine has an abrupt shutdown ($t>t_{\rm D}$), the flux sharply drops, as shown in Figure~\ref{fig:Lph} (b) and (d) for unsaturated and saturated acceleration regime where $t_{\rm D}$=0.01~s, 0.1~s, 1~s and 10~s respectively. Moreover, the observed light curves of the GRBs show relatively slow change in luminosity, unlike the steep rise and fall (almost within $10^{-3}$ s), thus, a continuous wind with variable luminosity is used to simulate a GRB pulse.

\section{HIC of photospheric emissions in different regimes}\label{sec:emissions}
There are three regimes discussed in the following analysis:  non-saturated emission where the regime of $R_{\rm s}>R_{\rm ph}$ is dominant for all over the wind profile; saturated emission where the regime of $R_{\rm s}\leq R_{\rm ph}$ works all over the wind profile; the third regime is emphasized where emissions are from different regimes, and defined as intermediate photospheres.

\subsection{The Non-saturated Emission}\label{sec:caseI}
In the unsaturated regime, the temperature of the photosphere, $T_{\rm ph}$ can be expressed as
 \begin{equation}\label{T2}
T_{\rm ph} \propto L_{\rm w}^{1/4},
\end{equation}
 given $E_{\rm p} \propto kT_{\rm ph}$ and $F\propto L_{\rm w}$, we have $\kappa=1/4$. Most of the prompt emission is from within $5/\Gamma_0$ around the line of sight independent of the opening angle. Note that this works if $L_{\rm w}$ is angle-independent. The case of an angle-dependent $L_{\rm w}$ will be shown in Section~\ref{sec:caseIV}. 
 
 \subsection{The Saturated Emission}\label{sec:caseIII}
 In the saturated emissions, $T_{\rm ph}$ can be expressed as
 \begin{eqnarray}
T_{\text{ph}}(\theta ) &\propto &L_{w}^{1/4}r_{0}^{-1/2}[R_{\text{ph}%
}(\theta )/R_{s}(\theta )]^{-2/3}  \notag \\
&\propto &L_{w}^{1/4}r_{0}^{-1/2}\{[L_{w}/\Gamma ^{3}(\theta )]/[\Gamma
(\theta )r_{0}]\}^{-2/3}  \notag \\
&\propto &L_{w}^{-5/12}r_{0}^{1/6}\Gamma ^{8/3}(\theta ).
\end{eqnarray}
 $\kappa$ is a negative value, and changes from -5/12 due to the jet structure. The simulation will be given in Section~\ref{sec:caseII} in a GRB pulse around the peak luminosity.
 
 \subsection{Intermediate Photospheres}\label{sec:caseII_IV}
In this analysis, we always firstly consider the luminosity could be approximated as angle-independent ($L(\theta)$ is taken as $L_{0}$, or denoted as $q=0$\footnote{Here in the following analysis, $q=0$ represents $L(\theta)=L_{0}$, rather than $L(\theta)=L_{0}/\sqrt{2}$ in Equation(\ref{eq:lumi_theta}). }) for simplicity. Assume the regime of $R_{\rm s}>R_{\rm ph}$ works in lower latitude, given that $R_{\rm s}=\eta(\theta)r_{0}$ monotonically decreases with $\theta$, while $R_{\rm ph}$ in Equation~(\ref{Rdcp3}) monotonically increases with $\theta$, it may turn to the regime of $R_{\rm s}\leq R_{\rm ph}$ in higher latitude when $\eta(\theta)\leq(\frac{Y}{\Gamma_{r_ 0}^2}\times\frac{\sigma_{\text{T}}}{m_{\text{p}}c}\frac{L}{4\pi c^2 r_0)})^{1/4}$. Therefore, unsaturated and saturated regime may both work in lower and higher latitude of the jet, and this case is denoted as intermediate photosphere~I in the following analysis. 

 Otherwise, if assuming saturated emission in lower latitude and a structured luminosity with a large enough $q$, $R_{\rm ph}$ may decrease more rapidly than $R_{\rm s}$ with increasing $\theta$, $R_{\rm s}>R_{\rm ph}$ may work in higher latitude if $\theta$ increases to that satisfies $\frac{L(\theta)}{(1+\beta)\beta\eta(\theta)^4}<4\pi c^2 r_0/\frac{\sigma_{\rm T}}{m_{\rm p}c}$. This case is defined as intermediate photosphere~II.   

In intermediate photosphere~I or II, there exists a critical value $\theta_{\rm cri}$, where the regime changes for $\theta \geq \theta_{\rm cri}$. Take intermediate photosphere~I as an example, at $\theta=\theta_{\rm cri}$, $R_{\rm ph}$ in Equation~(\ref{Rdcp3}) should be equal to that in Equation~(\ref{Rdcp}), where $R_{\rm ph}=R_{\rm s}$. The Equation~(4) in \cite{Meszaros_2000} shows the saturated  acceleration regime, $R^{>}_{\rm ph}/r_0=\frac{L_{\rm w}\sigma_{\rm T}Y}{4\pi r_0 m_{\rm p} c^3\eta^{3}}$, comparing to Equation~(\ref{Rdcp}), we have $R_{\rm ph}$ in the regime of $R_{\rm s}\gtrsim R_{\rm ph}$, $ R_{\text{ph}}=(\frac{\sigma _{\text{T}}}{2m_{\text{p}}c}%
\frac{d\dot{M}}{d\Omega }r_{0}^{2}) ^{1/3}$.

\subsubsection{intermediate photosphere~I: $R_{\rm s}>R_{\rm {ph}}$ and $R_{\rm s} \leq R_{\rm{ph}}$ in lower and higher latitude respectively}\label{sec:caseII}
 \begin{figure*}
\begin{center}
 \centering
  \includegraphics[width=\columnwidth]{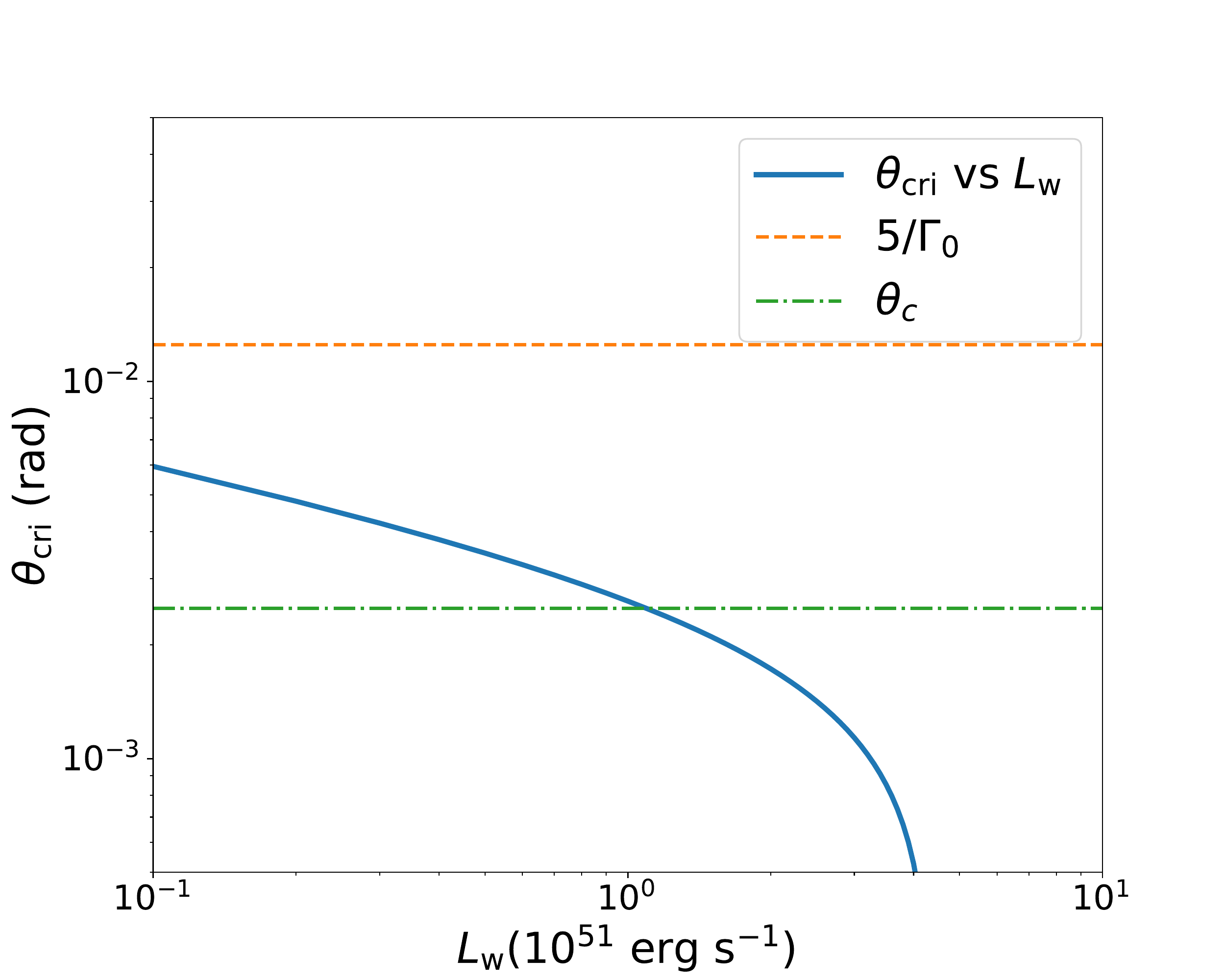}\put(-200,160){(a)}
  \includegraphics[width=\columnwidth]{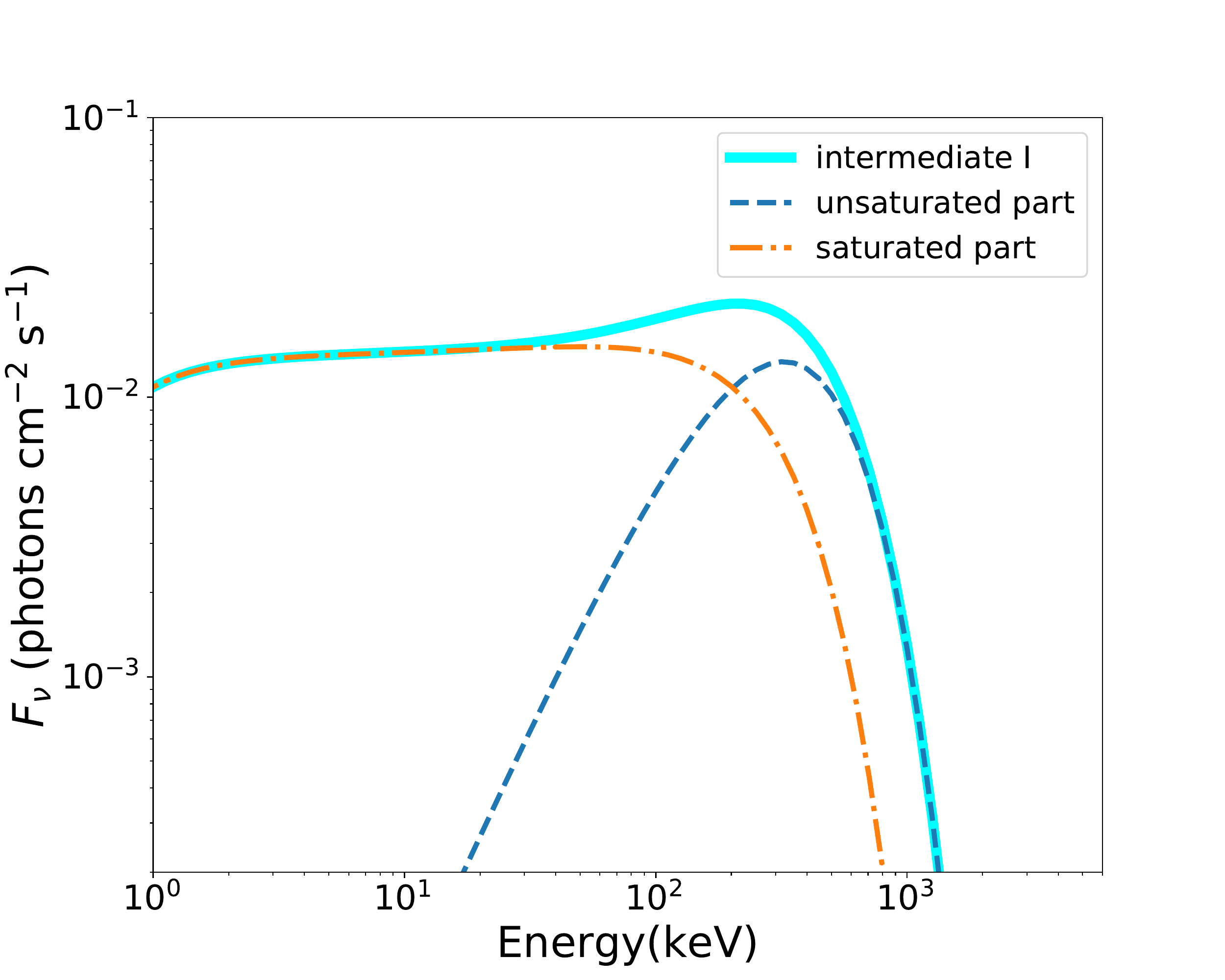}\put(-200,160){(b)}
 \caption{(a) The relation between $\theta_{\rm cri}$ and $L_{\rm w}$ in intermediate photosphere I is shown, and denoted as the blue line. The orange dashed line denotes the value of $5/\Gamma_0$ and the green dot-dashed line denotes the value of $\theta_{\rm c}$. 
  (b) $F_{\nu}$ spectrum of intermediate photosphere of $1.0L_{\rm w,51}$ is denoted by the cyan thick line. The blue dashed line and orange dot-dashed line denote the part from unsaturated and saturated emissions.
 \label{fig:theta_cri_spectra_II_III}}
\end{center}
\end{figure*} 
Figure~\ref{fig:theta_cri_spectra_II_III}~(a) shows how $\theta_{\rm cri}$ varies with $L_{\rm w}$ in intermediate photosphere~I.
 The parameters in the simulation of the continuous wind are listed as bellow: $L_{\rm w}$ ranges from $0.1$ to $10.0L_{\rm{w}, 51}$ ( $L_{\rm{w}, 51}=10^{51}$ erg s$^{-1}$. Here and below we always use CGS units\footnote{the convention $Q = 10^{n}Q_{n}$ is adopted for CGS units.}.); $r_0$= $10^8$~cm, $\Gamma_0=400$\footnote{From the fireball samples in \cite{Pe_er_2015}, $r_0$ spans a wide range from $10^{6.5}$ to $10^{9.5}$ cm with the mean value of $10^{8}$ cm, and $\Gamma$ from $10^{2}$ to $10^{3}$ with the mean value of 370, here we take $\Gamma=400$ as an approximate moderate value.}, $p=1$ , $\Gamma_0\theta_{\rm c}=1$, and $\theta_{\rm v}=0$. In this analysis, we always take $d_L=4.85\times10^{28}$~cm which corresponds with z=2\footnote{ $d_{\rm L}=4.85\times10^{28}$~cm and z=2 corresponds to the peak of the GRB formation rate according to \cite{2016A&A...587A..40P}}.
  In Figure~\ref{fig:theta_cri_spectra_II_III}~(a), $\theta_{\rm {cri}}$ is smaller than the value of $5/\Gamma_0$, which means the saturated emission contributes to the prompt part of GRB. For the samples with $1.0L_{\rm{w}, 51}<L_{\rm w}<4.0 L_{\rm{w}, 51}$, $\theta_{\rm {cri}}$ becomes less than $\theta_{\rm c}$ and decreases with increasing $L_{\rm{w}}$, which means the contribution from the saturated emission increases. Note that above 4.0$L_{\rm{w}, 51}$, the emission from the regime of $R_{\rm ph}\geq R_{\rm s}$ is dominant, and it is expected to nearly behave like the saturated emission. In this case, $R_{\rm dcp}$ takes the form of Equation~(\ref{Rdcp3}) because $R_{\rm ph}\lesssim R_{\rm s}$ works, and above 4.0$L_{\rm{w}, 51}$, the emission from $R_{\rm ph}\geq R_{\rm s}$ is dominant and it turns to the saturated regime.
 
 In Figure~\ref{fig:theta_cri_spectra_II_III}~(b), the $F_{\nu}$ spectrum of the intermediate photospheric emission of $1.0L_{w,51}$ is shown in the cyan thick line. The contributions from unsaturated and saturated emissions are denoted by the blue dashed and orange dot-dashed lines. $\alpha$ of the spectrum of intermediate photosphere is smaller than that from the unsaturated regime only. Thus, it could be expected that $\alpha$ decreases with the contribution from the saturated regime increasing with $L_{\rm w}$. 
\begin{figure*}
\begin{center}
 \centering
  \includegraphics[width=\columnwidth]{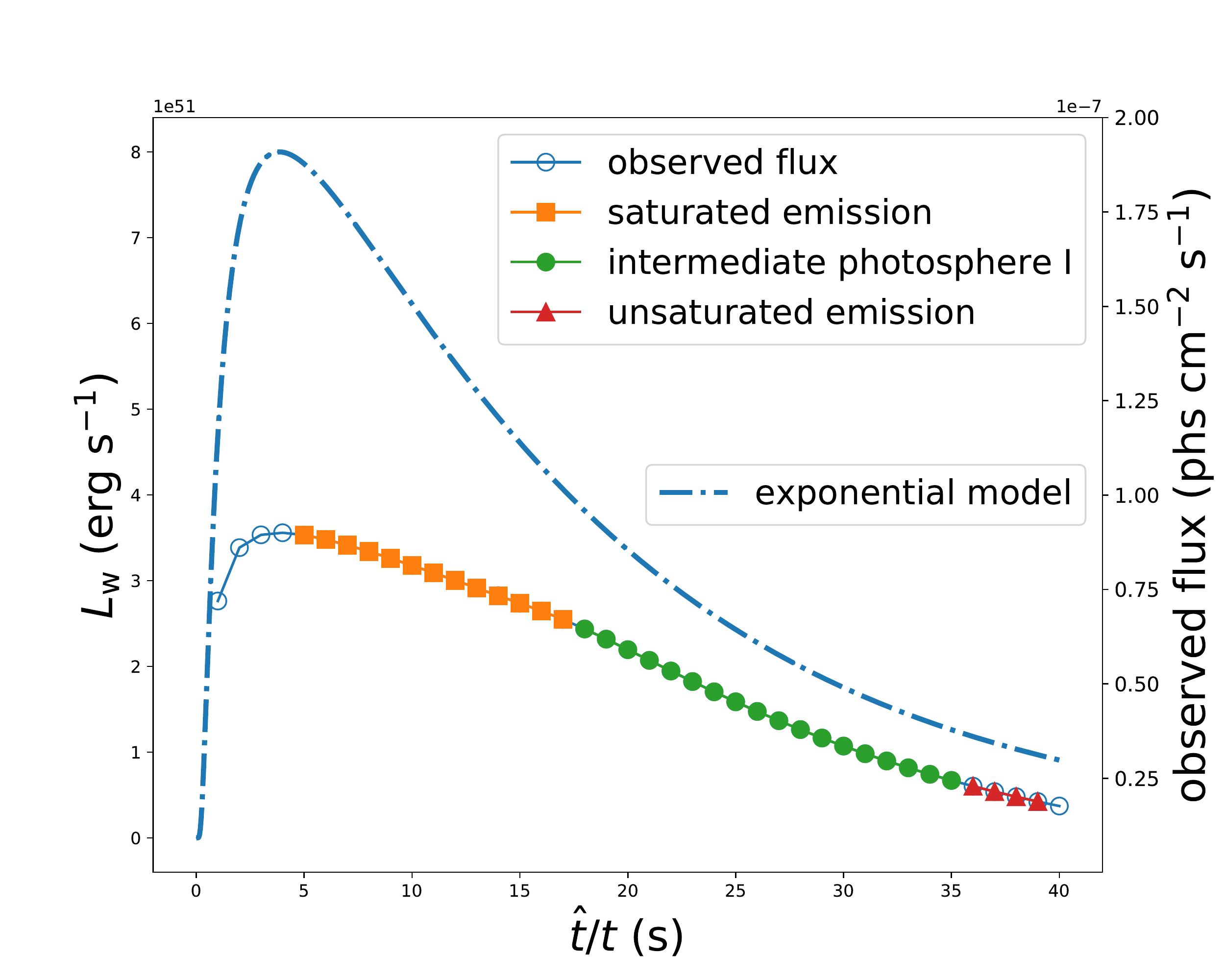}\put(-210,160){(a)}
  \includegraphics[width=\columnwidth]{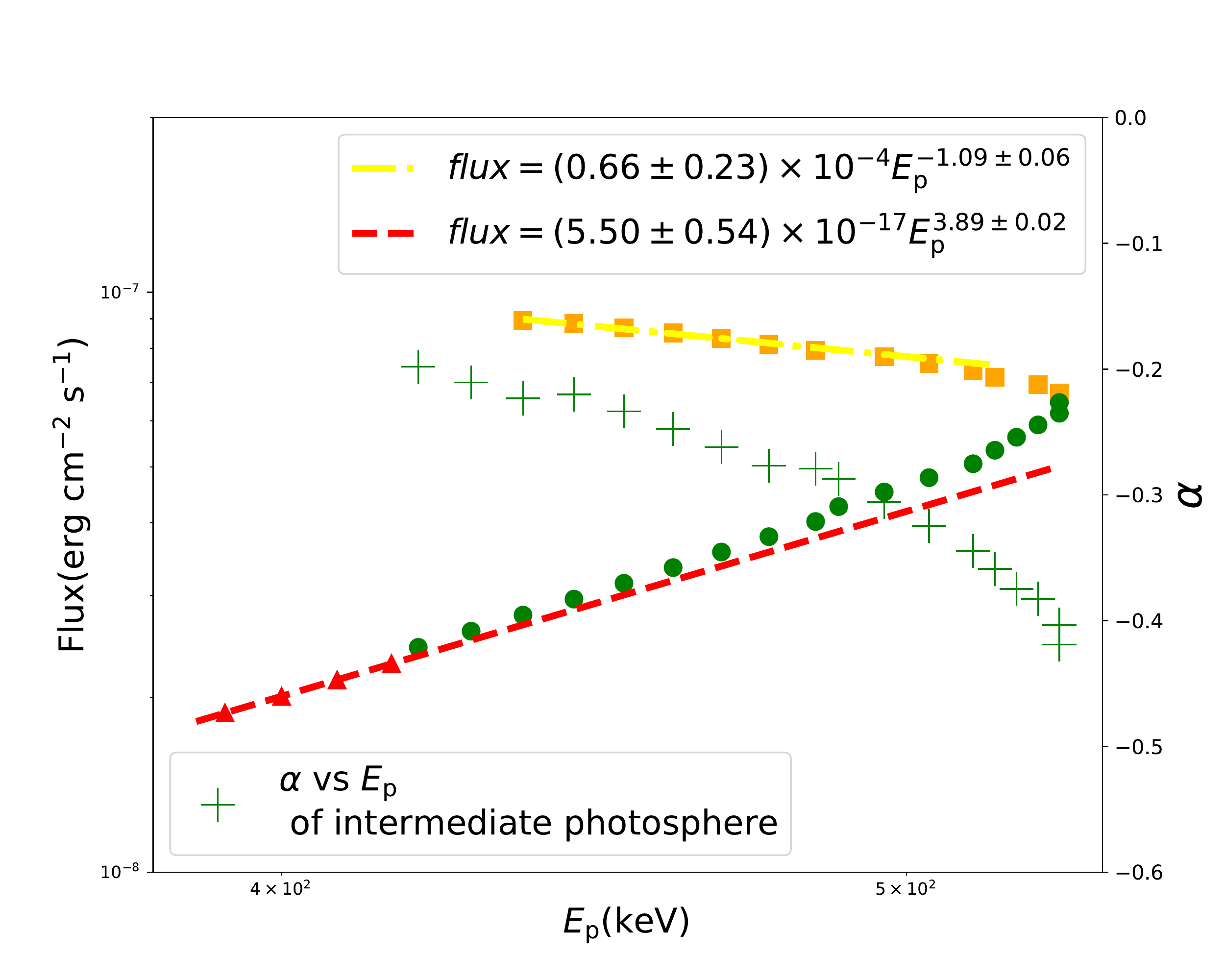}\put(-200,160){(b)}
\caption{ (a) The shapes of the exponential GRB pulse model is denoted by the dot-dashed blue line. The orange squares, green dots and red triangles denote the observed flux of emissions from the saturated, intermediate photosphere I and unsaturated regimes respectively, and the same below. (b) $F-E_{\rm p}$ and $\alpha-E_{\rm p}$ of different regimes.  \label{fig:Lw_Flux_Ep_alpha}}
\end{center}
\end{figure*}

  To mimic a GRB pulse, an exponential model\footnote{$L_{\rm w}(\hat{t})$ of exponential model is described as $L_{\rm w}(\hat{t} >\hat{t}_{\rm s})=L_{\rm w,p}\times \exp \left[ 2\left( \tau
_{1}/\tau _{2}\right) ^{1/2}\right]  \times \exp \left( -\frac{\tau _{1}}{\hat{t}-\hat{t%
}_{\rm s}}-\frac{\hat{t}-\hat{t}_{\rm s}}{\tau _{2}}\right)$, where $\hat{t}_{\rm s}$ is the start time, $\tau_{1}$ and $\tau _{2}$ are respectively the characteristic time scales indicating the rise and decay
periods, $L_{w,p}$ is the peak of luminosity at $\hat{t}_{p}$, and $%
\hat{t}_{p}=\hat{t}_{s}+\left( \tau _{1}\cdot \tau _{2}\right) ^{1/2}$. }~\citep{2005Long} is utilized  with ($\tau_1$, $\tau_2$, $\hat{t}_{p}$, $L_{\rm w, p}$)=(1, 15, -0.02, $0.8L_{\rm w, 52}$). In Figure~\ref{fig:Lw_Flux_Ep_alpha}~(a), the orange squares, green dots and red triangles denote the flux of emissions from the saturated, the intermediate photosphere and the unsaturated regime respectively.
 As shown in Figure~\ref{fig:Lw_Flux_Ep_alpha}~(b), $F-E_{\rm p}$ of unsaturated emissions gives 
$index\sim4$, $F-E_{\rm p}$ of intermediate photosphere I gives a larger $index$ ($\kappa<1/4$), and then turns to have a negative $\kappa\sim-1$ when the saturated emission becomes dominant with increasing $L_{\rm w }$. The values of $\alpha$ of intermediate photosphere I decrease from -0.2 to -0.4, which means that there exists anti-correlation in $\alpha-E_{\rm p}$ in this regime, which is consistent with our prediction above.

\subsubsection{intermediate photosphere~II:$R_{\rm s}\leq R_{\rm ph}$ and $R_{\rm s}>R_{\rm{ph}}$ in low- and high-latitude respectively}\label{sec:caseIV}
\begin{figure*}
  \includegraphics[width=\columnwidth]{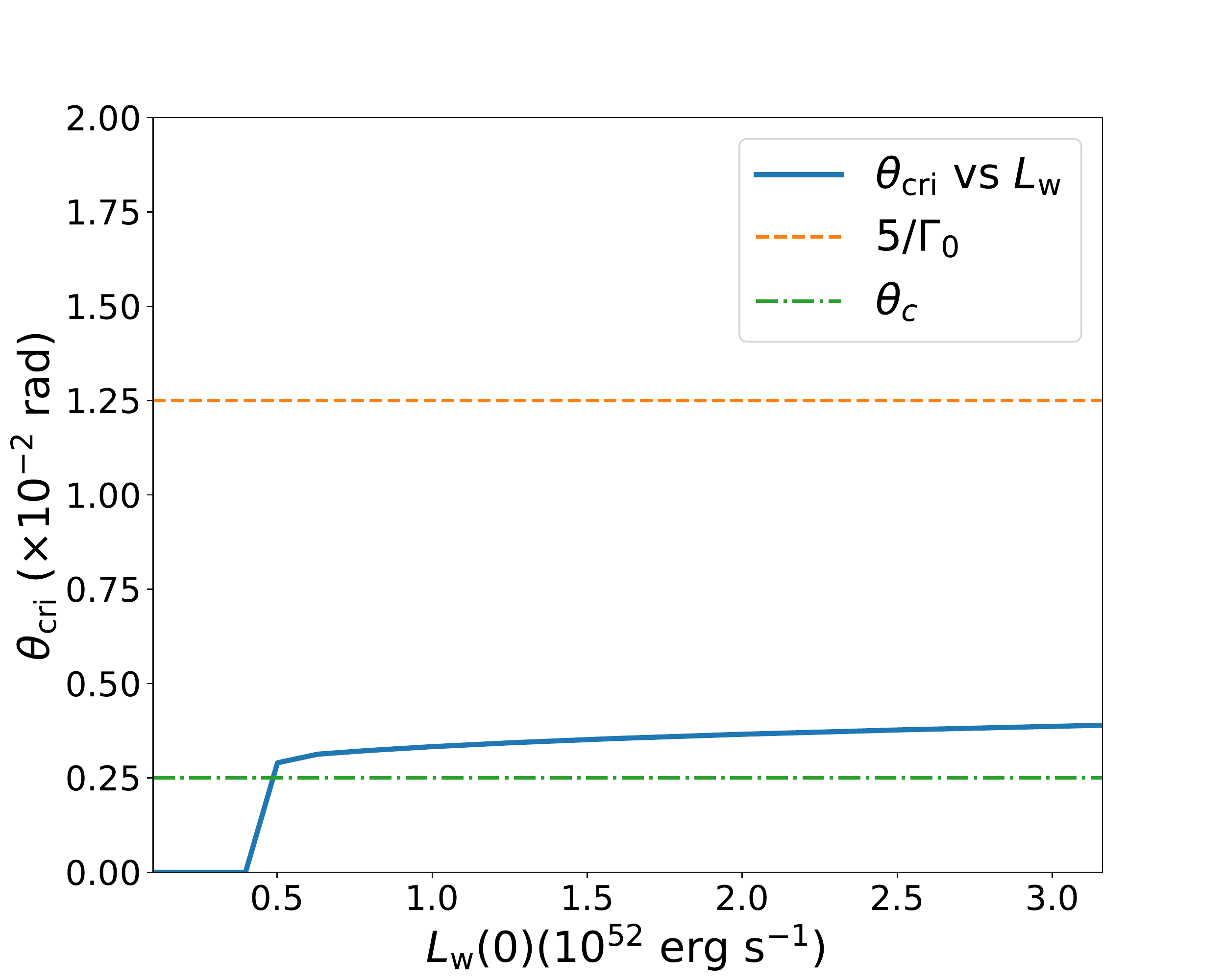} \put(-200,160){(a)}
  \includegraphics[width=\columnwidth]{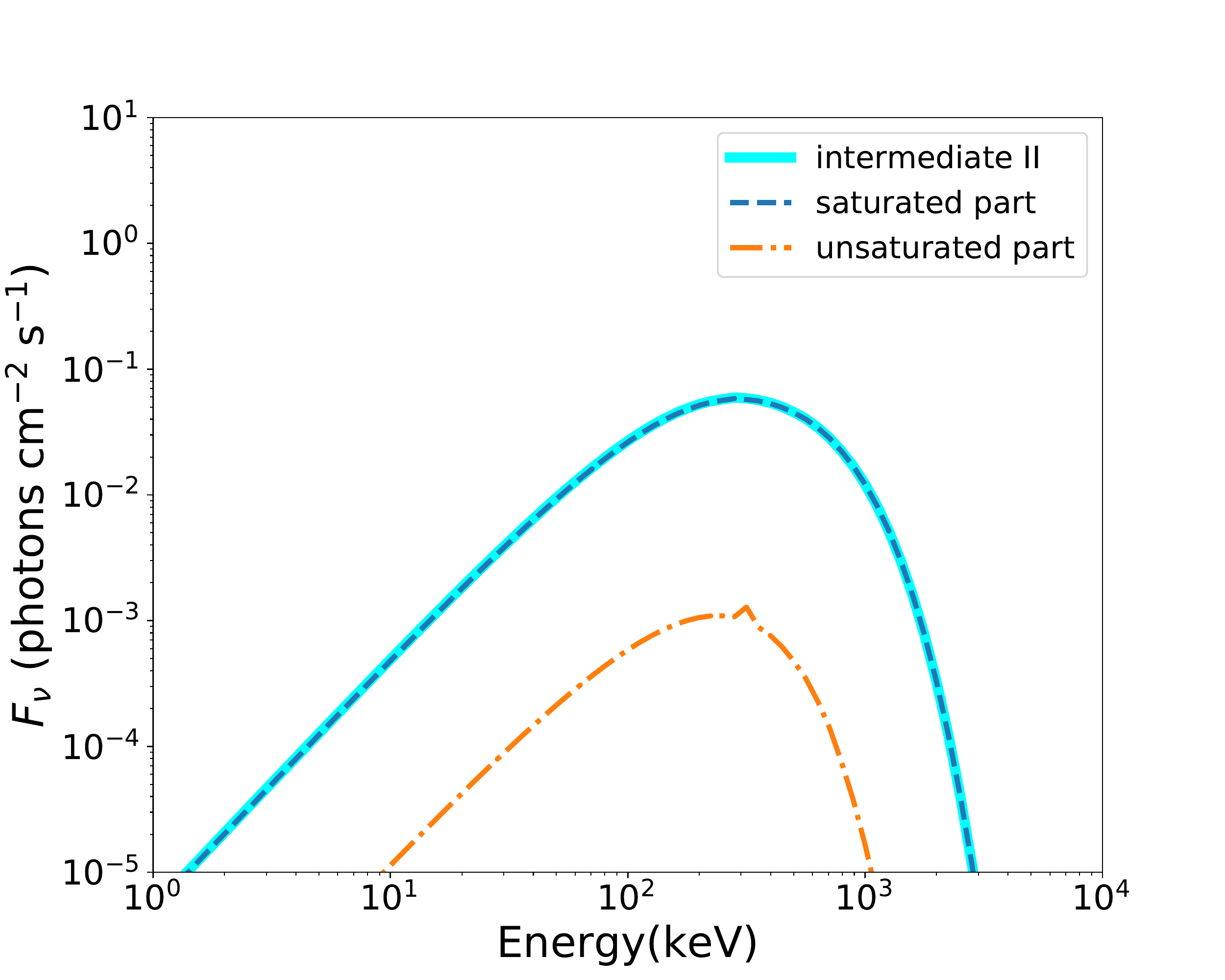}\put(-200,160){(b)}
\caption{ (a) The relation between $\theta_{\rm cri}$ and $L_{\rm w}$ in intermediate photosphere II. 
  (b) $F_{\nu}$ spectrum of intermediate photosphere II of $0.5L_{w,52}(0)$ is denoted by the cyan thick line. The blue dashed line and orange dot-dashed line denote the part from saturated and unsaturated parts.\label{fig:caseIV_thetacri_spectra}}
\end{figure*}

\begin{figure*}
\includegraphics[width=\columnwidth]{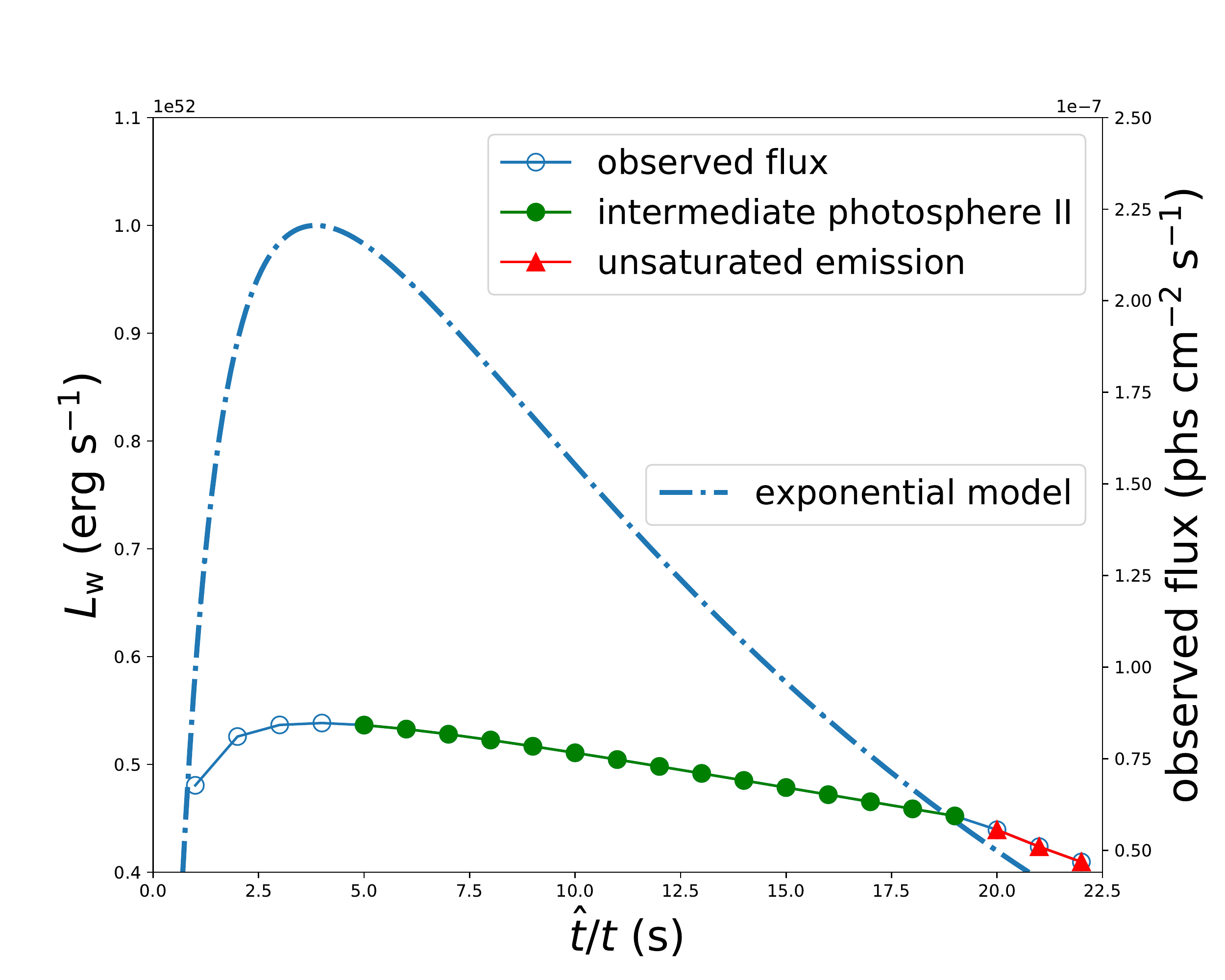}
  \put(-200,160){(a)}
   \includegraphics[width=\columnwidth]{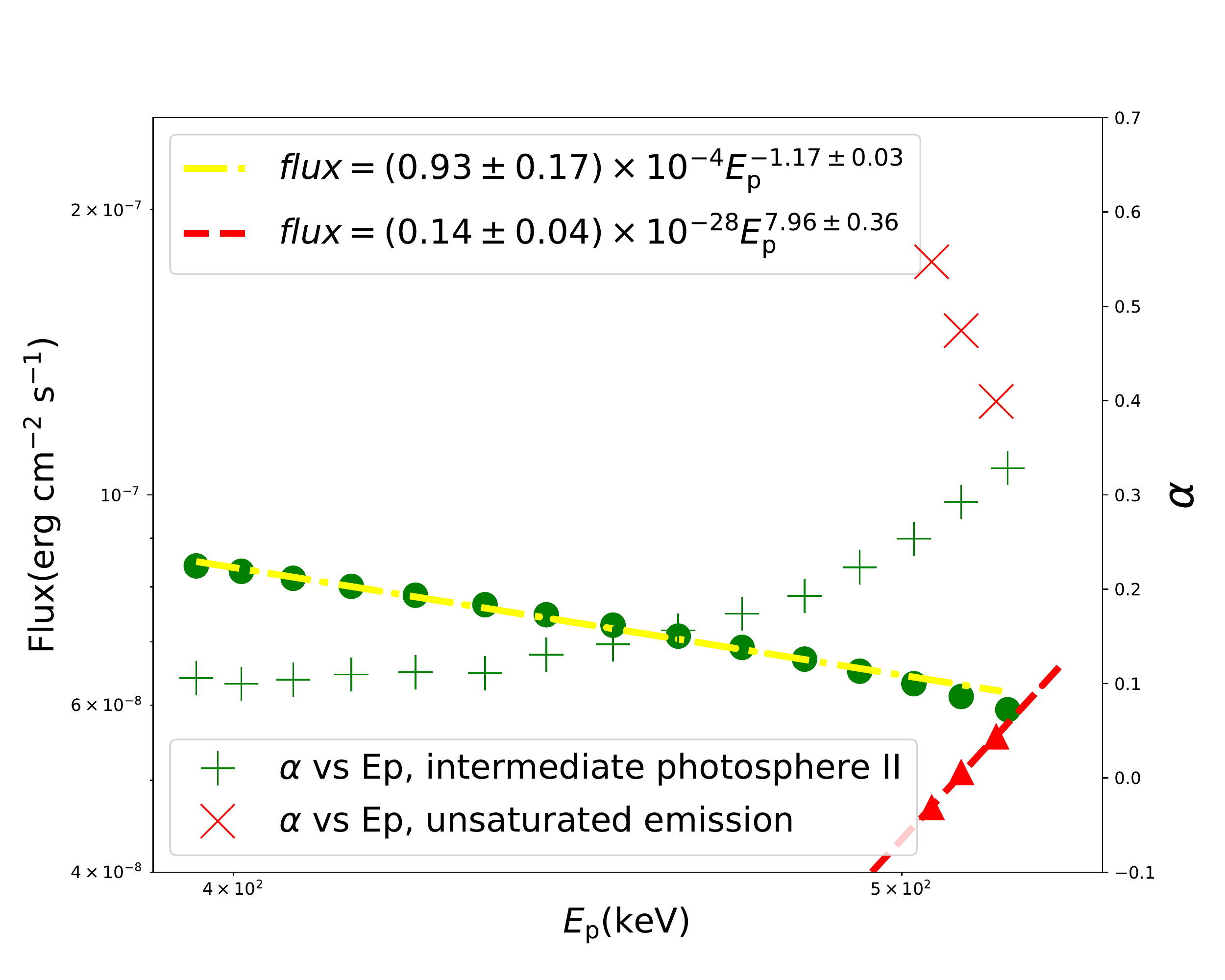} \put(-40,160){(b)}
\caption{ (a)The shape of exponential GRB pulse model and the observed flux with angle-dependent $L_{\rm w, p}(0)=L_{\rm w, 52}$. The green dots and red triangles denote the observed flux of emissions from intermediate photosphere II and unsaturated regimes respectively, and the same below. (b) $F-E_{\rm p}$ and $\alpha-E_{\rm p}$ of different regimes. \label{fig:Lw_Flux_Ep_alpha2}}
\end{figure*}
$\theta_{\rm {cri}}$ of intermediate photosphere~II are shown in Figure~\ref{fig:caseIV_thetacri_spectra}~(a) of $L_{\rm w}(0)$ ranging from $0.5$ to 3.2$L_{\rm{w}, 52}$ with $q=10$ and $\theta_{\rm c,L}$=$\theta_{\rm c}$. The other parameters are the same as those used in intermediate photosphere~I. As shown in  Figure~\ref{fig:caseIV_thetacri_spectra}~(a), $\theta_{\rm {cri}}$ is less than $5/\Gamma_0$ for all over the selected range of $L_{\rm{w}}(0)$. Unsaturated emissions with $L_{\rm w}(0)$ ranging from $0.1$ to 0.4$L_{\rm{w}, 52}$ have $\theta_{\rm cri}=0$, are plotted as well for comparison. 
  From the $F_{\nu}$ spectrum of $0.5L_{w,52}(0)$ shown in Figure~\ref{fig:caseIV_thetacri_spectra}~(b), the contribution from the unsaturated emission is very small and even smaller with $L_{\rm w}(0)$ increasing. Therefore, the prompt emission of intermediate photosphere~II could be regarded to be dominant by the saturated emission.
   
   Figure~\ref{fig:Lw_Flux_Ep_alpha2}~(a) shows the flux of the simulated GRB pulse with $L_{\rm w,p}=1.0L_{\rm w,52}$, and the ranges of intermediate photosphere II and unsaturated emissions. As shown in  Figure~\ref{fig:Lw_Flux_Ep_alpha2}~(b), the emission of the intermediate photosphere II have an anti-correlation of $F-E_{\rm p}$ with $\kappa\sim-1$. 
  Besides, as shown in red triangles in Figure~\ref{fig:Lw_Flux_Ep_alpha2}~(b), we also find that in unsaturated emission with considering an angle-dependent luminosity, $index>4$ ($\kappa<1/4$).
  
  It seems that the angle-dependent luminosity structure could enhance the evolution of $\alpha-E_{\rm p}$. Anti-correlation is found in $\alpha-E_{\rm p}$ in red crosses which denote the unsaturated emissions in Figure~\ref{fig:Lw_Flux_Ep_alpha2}~(b), while positive correlation between $\alpha-E_{\rm p}$ in green plus markers is found in the saturated emissions.

\section{Discussion AND CONCLUSION}\label{sec:discuss}
\subsection{HIC and trend of $F-E_{\rm p}$ }
 $index=4$ or $\kappa=1/4$ could be taken as a characteristic value of HIC for unsaturated emission.
   In the intermediate photosphere, $\kappa$ varies from 1/4 to an even smaller value, and then turns to negative as that in the saturated emission.  This could be explained as follows: as the luminosity of the outflow increases, the emissions tend to be dominant by the saturated regime; the flux increases while $E_{\rm p}$ does not, due to the anti-correlation in $F-E_{\rm p}$ of the saturated regime, therefore, $\kappa$ falls from 1/4 to negative continuously, and never reach up to the value of $1/3$, $1/2$ or even larger. In the decay phase of the second (main) pulse ($t>30$ s) in GRB 081221A, a correlation of $F\varpropto (kT)^{4.1}$ is given~\citep{Hou_2018}, where $kT$ is the temperature of blackbody model and $kT \varpropto E_{\rm p}$.  
   
The similar trend of $F-E_{\rm p}$ could be generalized to the emission of the photosphere from a structured jet in the case of a hybrid relativistic outflow. As shown in Table 1 in \cite{Gao_2015}, for the case of no magnetic dissipation, $kT_{\rm ob}\varpropto L_{\rm w}^{\kappa_1}$, with $\kappa_1$ of 1/4, -1/60, -5/12, $F_{\rm BB}\varpropto L_{\rm w}^{\kappa_2}$ with $\kappa_2$ of 1, 11/15, 1/3 from unsaturated to saturated regime, where $k_{\rm B}T_{\rm ob}$ and $F_{\rm BB}$ denote the observed temperature of the thermal component and the thermal flux. With increasing $L_{\rm w}$, $R_{\rm ph}$ increases and emissions from saturated acceleration regime become dominant, which behaves similar to that of pure hot fireball component. $\kappa=\kappa_1/\kappa_2$ in unsaturated emissions is equal to 1/4, which means $\kappa$ never reaches up to a larger value. For the case of considering magnetic dissipations, if the emission is fully thermalized, the result is similar. 
 Therefore, photospheric emissions can not contribute to the peak distribution at $\kappa\sim1/2$ in the observation shown in \cite{2000On} and \cite{Lu:2012pf}. 
   
\begin{figure} 
\begin{center}
 \centering
  \includegraphics[width=\columnwidth]{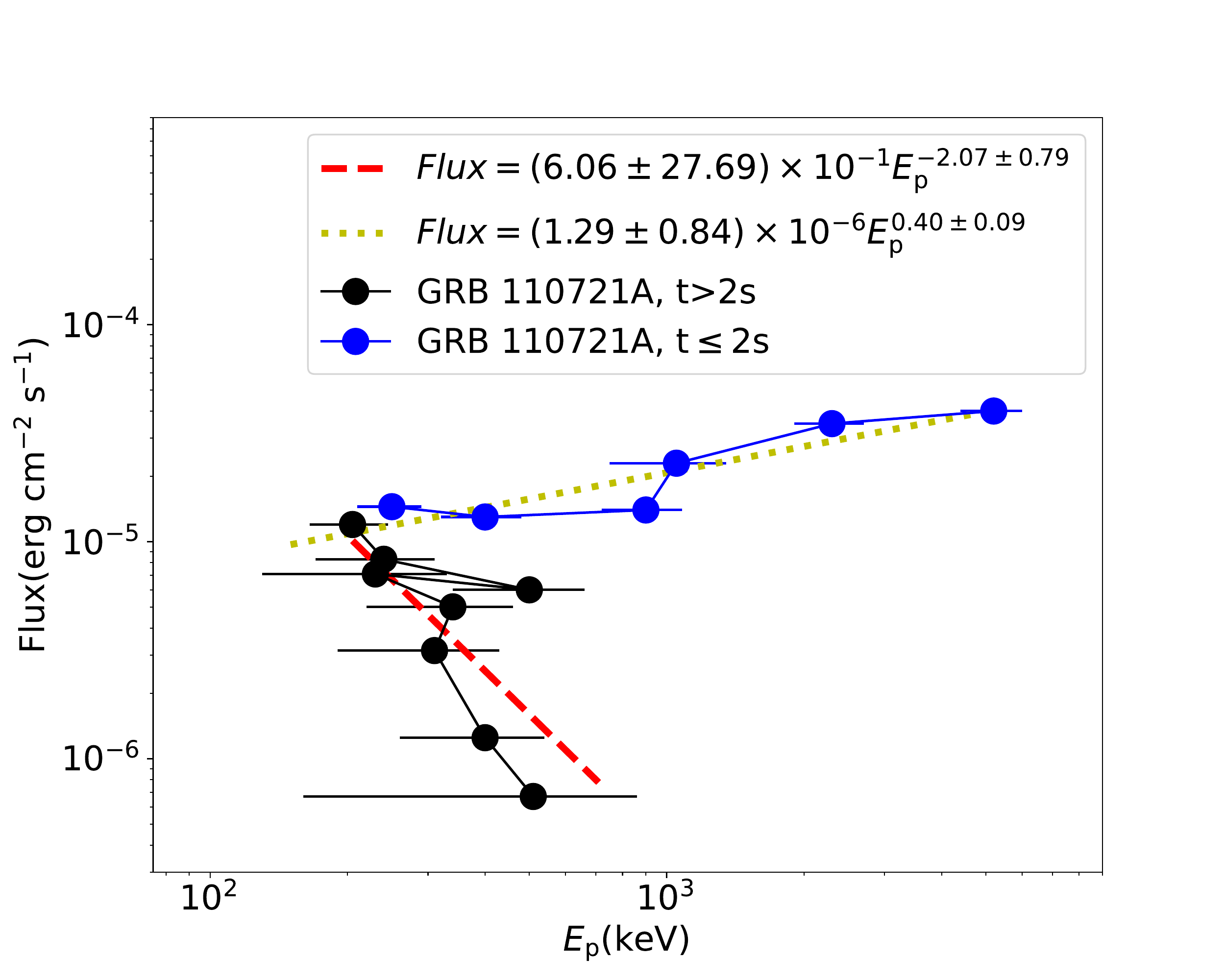}\\
\caption{$F-E_{\rm p}$ of GRB 110721A is adapted from the fourteenth panel of Figure 2 on Page 6 in \protect\cite{Lu:2012pf} and $F-E_{\rm p}$ per dot is extracted from fitting to time-resolved spectra with BAND model. The blue dots are from $t\leq2$ s, while the black ones represent those from $t>2$ s. The red dashed and yellow dotted lines represent the fitted curve of the two trends. \label{fig:GRB110721A}}
\end{center}
\end{figure}

 HIC of photospheric emissions from saturated acceleration regimes is anti-correlation. It is observed in $F-E_{\rm p}$ of GRB110721A. As shown in Figure~\ref{fig:GRB110721A}, the two trends are fitted respectively. A correlation with negative $index=-2.07\pm0.79$ is fitted from $F-E_{\rm p}$ of $t>2$ s with $E_{\rm p}$ from 200 keV to 500 keV in the second pulse. $index$ of HIC in this range is consistent with that of photospheric emission from saturated acceleration regimes. As discussed in \cite{stt863} and \cite{Gao_2015}, the largest proportion of thermal flux is reached after 2 s. The positive $index\sim0.5$ is obtained in $F-E_{\rm p}$ of $t\leq 2$ s mainly from the first pulse, with $E_{\rm p}$ from 200 keV to 5 MeV. The positive $index<4$ implies that the emission from the non-thermal component are dominant in this range, which is roughly consistent with the analysis of \cite{stt863} and \cite{Gao_2015}.
 
\subsection{$\alpha-E_{\rm p}$ evolution in a GRB pulse } 
 Anti-correlation in $\alpha-E_{\rm p}$ could be naturally reproduced by intermediate photosphere I. Besides, the angle-dependent luminosity could also enhance the evolution of $\alpha-E_{\rm p}$, reproduce positive- and anti-correlation in the saturated and unsaturated emissions.
 
 A positive correlation in $F-E_{\rm p}$ and anti-correlation in $\alpha-E_{\rm p}$ in both two pulses of GRB 120728A are shown in Figure A7 and A6 in \cite{Li_2021}. Moreover, $\alpha$ is large, nearly up to about 1, which implies the prompt emission may be dominant by the photospheric emission in the intermediate photosphere I, however, we need further study on these similar GRBs.
 
 \subsection{Is the intermediate photosphere  unusual?}
\begin{figure}
\begin{center}
 \centering
  \includegraphics[width=\columnwidth]{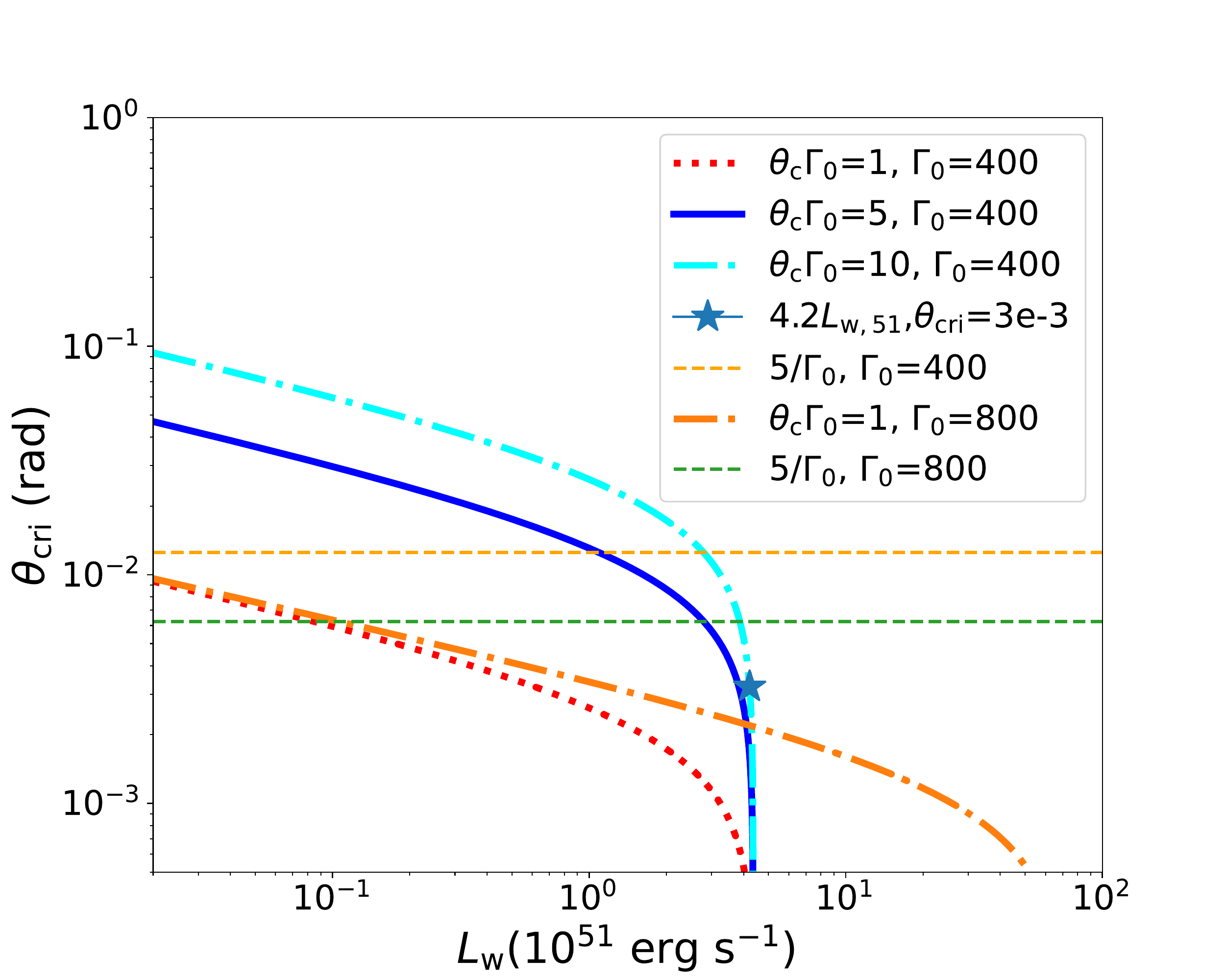}\put(-200,160){(a)}\\
  \includegraphics[width=\columnwidth]{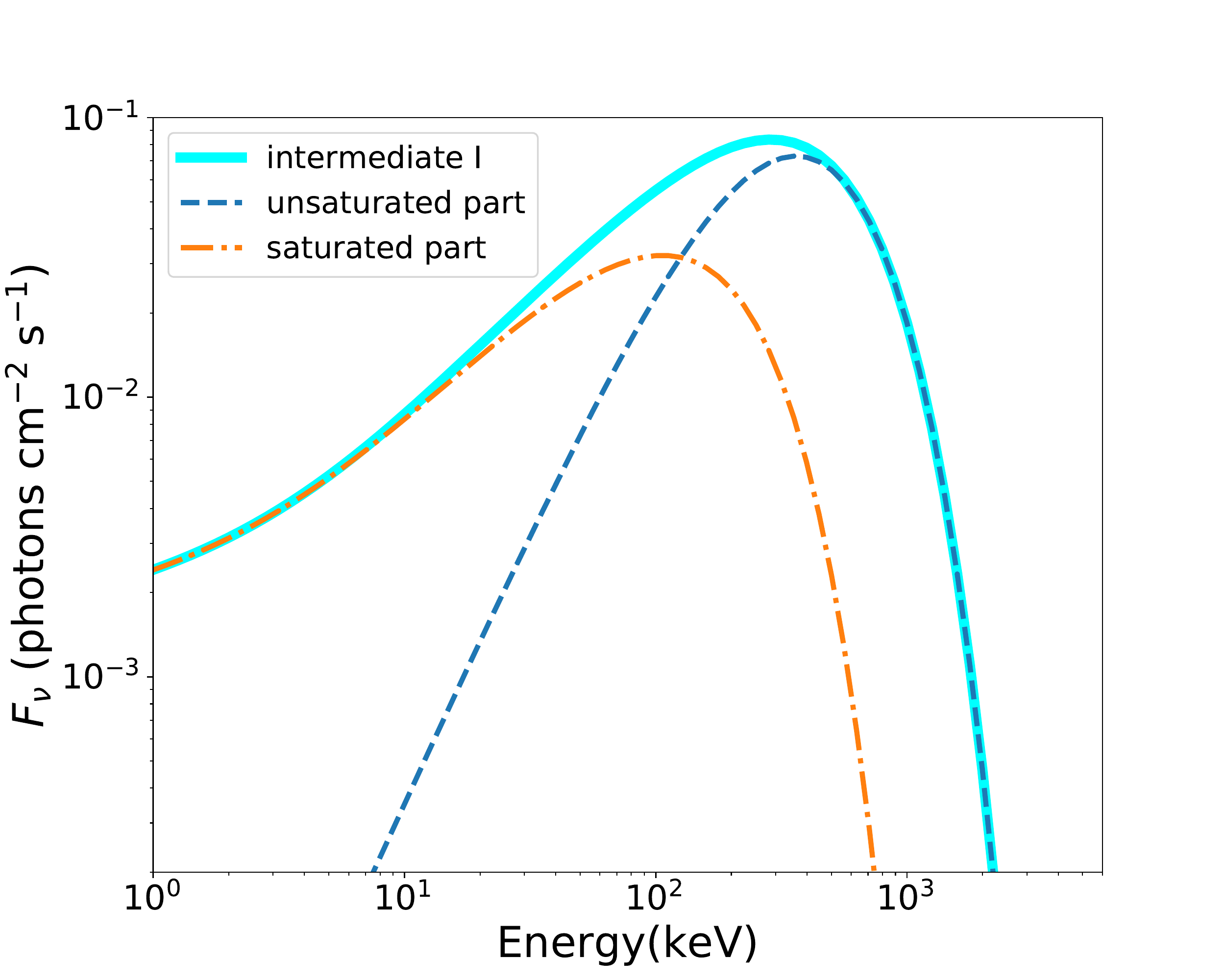}\put(-200,100){(b) $\theta_{\rm c}\Gamma_0=10$ and $\Gamma_0=400$ }
\caption{ (a) The $\theta_{\rm cri}$ of $\theta_{\rm c}\Gamma_0=1$, 5, 10 with $\Gamma_0=400$ denoted by red dotted, blue solid and cyan dot-dashed lines. The star denotes $(L_{\rm w},\theta_{\rm cri} )=(4.2L_{\rm w,51}, 3\rm e-3)$ in the case of $\theta_{\rm c}\Gamma_0=10$ with $\Gamma_0=400$ on the cyan dot-dashed line. The orange dot-dashed line denotes $\theta_{\rm cri}$ with $\theta_{\rm c}\Gamma_0=1$ and $\Gamma_0=800$. (b) $F_{\nu}$ spectrum of the intermediate photosphere of the star shown in (a), is denoted by the cyan thick line. The blue dashed line and orange dot-dashed line denote the part from unsaturated and saturated parts. \label{fig:theta_cri_diffwidth}}
\end{center}
\end{figure}

 As shown in Figure~\ref{fig:theta_cri_diffwidth}~(a), given a moderate $r_{0}$, $\Gamma_0$ and a assumption of a narrow jet ($\theta_{\rm c}\Gamma_0\lesssim$few), the range of $L_{\rm w}$ of intermediate photosphere is $10^{51-52}$ erg s$^{-1}$, which is a lower to moderate luminosity. If with a larger $\Gamma_0$, $L_{\rm w}$ at $\theta_{\rm cri}=0$ will be larger and the range will be broader, e.g. $L_{\rm w}$ reaches up to $\sim10^{53}$ erg s$^{-1}$ with $\Gamma_0=800$ denoted by the orange dot-dashed line. For a broader jet, e.g. $\theta_{\rm c}\Gamma_0=10$ with $\Gamma_0=400$ denoted by the cyan dot-dashed line, the range of $L_{\rm w}$ within $0<\theta_{\rm cri}< 5/\Gamma_0$ is narrower than that of $\theta_{\rm c}\Gamma_0=1$, which implies that, the intermediate photosphere mostly happens in a narrow jet rather that a broader one. However, even in a broad jet labeled by a star of on the line of $\theta_{\rm c}\Gamma_0=10$ with $\Gamma_0=400$ in Figure~\ref{fig:theta_cri_diffwidth}~(a), the saturated part still changes the spectrum within the detection range ($>8$ keV) and causes a smaller $\alpha$ as shown in the $ F_{\nu}$ spectrum of $(L_{\rm w},\theta_{\rm cri} )=(4.2L_{\rm w,51}, 3\rm{e}-3)$ in Figure~\ref{fig:theta_cri_diffwidth}~(b). Thus, the intermediate photospheres are not unusual in narrow jet, and for a broad jet with some specific luminosity, it may exist and should be considered.

\section{Summary}\label{sec:sum}
In this analysis, we discuss the intermediate photosphere and find that $\kappa$ of HIC is less than 1/4, and the same conclusion could be generalized to the photospheric emission from a hybrid relativistic outflow without magnetic dissipations or that with sub-photospheric magnetic dissipations and completely thermalized. Compared with the distribution of observed $\kappa$ in \cite{2000On} and \cite{Lu:2012pf}, $\kappa$ from photospheric emissions in these cases are almost beyond three standard deviations. Standard synchrotron model~\citep{2002An} gives a relation of $E_{\rm p} \propto \Gamma_{\rm e} L^{1/2}_{\rm w}R^{-1} (1+z)^{-1}$, where $\Gamma_{\rm e}$ is the typical electron Lorentz factor in the emission region, and $R$ is the emission radius. It naively gives $\kappa=1/2$, although the dependent is non-trivial considering other factors (e.g. flux is related to the evolution of the strength of the shock during the shock crossing and the number of electrons that are shocked, and this is also discussed in  \cite{Lu:2012pf}). This may imply that contribution peaking at $\sim1/2$ in the distribution of observed $\kappa$ are mainly from the prompt emission of GRBs with synchrotron origin. Therefore, $\kappa$ may offer a criterion for the origin of the prompt emissions of GRBs. Besides, emissions from the intermediate photosphere could naturally reproduce anti-correlation of $\alpha-E_{\rm p}$ in a GRB pulse. 

In this paper, some factors are ignored: firstly, sideway diffusion effect of photons at certain angular distances could cause a smearing out effect on the temperature, and lead to a non-thermal spectrum
due to the inverse-Compton radiation for jets~\citep{2013A, Ito_2013}. However, it is not considered, and we assume the evolution of outflow in each angular fluid element is independent; 
secondly, besides magnetic dissipations, the energy dissipation in the area of moderate optical depth is proposed by \cite{2012MNRAS.422.3092G} and \cite{2013ApJ...764..157B}. If it is considered, the non-thermal spectrum above the peak energy would be formed~\citep{2012MNRAS.422.3092G}, and affects the measurement of HIC; thirdly, in the simulation of teh GRB pulse, $t_{\rm D}$ could be treated as a tiny equal time interval in which $L_{\rm w}$ could be regarded as a constant, thus, $t_{\rm D}$ could be taken as minimum variability timescales (MVT). \cite{2015THEENERGY} shows that there exists energy dependence of MVT, where MVT in higher energy band is smaller than that in lower energy band. If related to the central engine, the higher $L_{\rm w}$ with smaller $t_{\rm D}$ and lower ones with larger $t_{\rm D}$, we can conclude that $index$ should be lower than 4. However, there is not enough information about MVT of different $L_{\rm w}$ for the emission of the central engine; fourthly, some correlations are not considered, such as the bulk Lorentz factor and the isotropic luminosity \citep[$\Gamma-L_{\gamma,\rm iso}$,][]{2012LORENTZ}. However, a basic assumption in a GRB pulse in this paper is that $L_{\rm w}$ is the only variable of $\hat{t}$, while the other parameters ($\theta_{\rm c}$, $p$, $q$, $r_0$, $\theta_{\rm v}$, $\Gamma_{0}$) remain constant. Therefore, correlations between the parameters are not considered.

\section*{Acknowledgements}
Xin-Ying Song thanks the support from Prof. Shao-Lin Xiong and Prof. Wen-Xi Peng during the work. We are very grateful for the comments and suggestions of the anonymous referees. In particular, we thank the GBM team for providing the GRB data that were used in this research.

\section*{Data Availability}
The data underlying this article will be shared on reasonable request to the corresponding author.



\bibliographystyle{mnras}
\bibliography{NDPandHICbib} 



\appendix
\section{The plots for impulsive injection and horizontal comparison results of $Case$~I}

\begin{figure*}
\begin{center}
 \centering
  \includegraphics[width=\columnwidth]{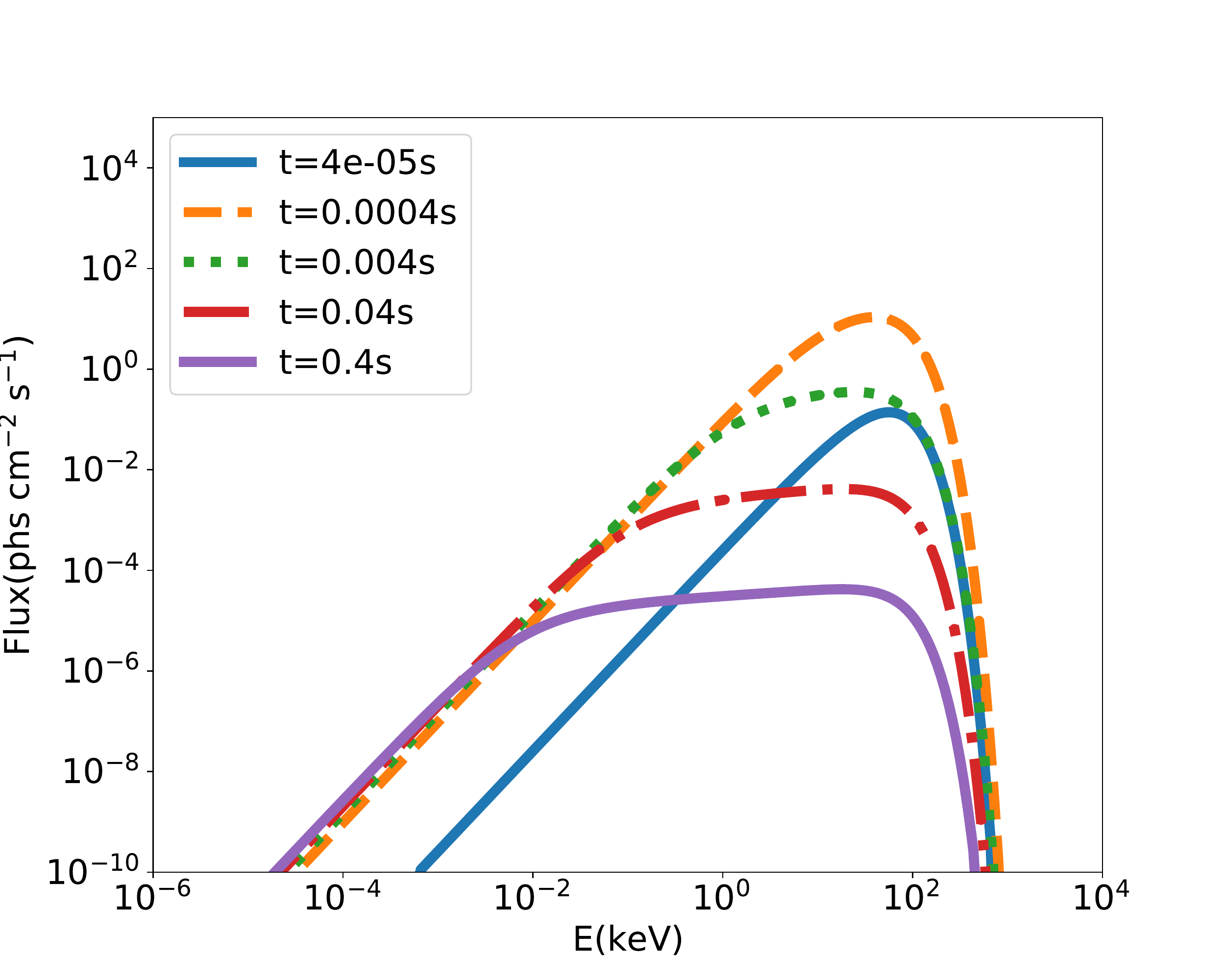}\put(-150,160){(a) the unsaturated emission}
  \includegraphics[width=\columnwidth]{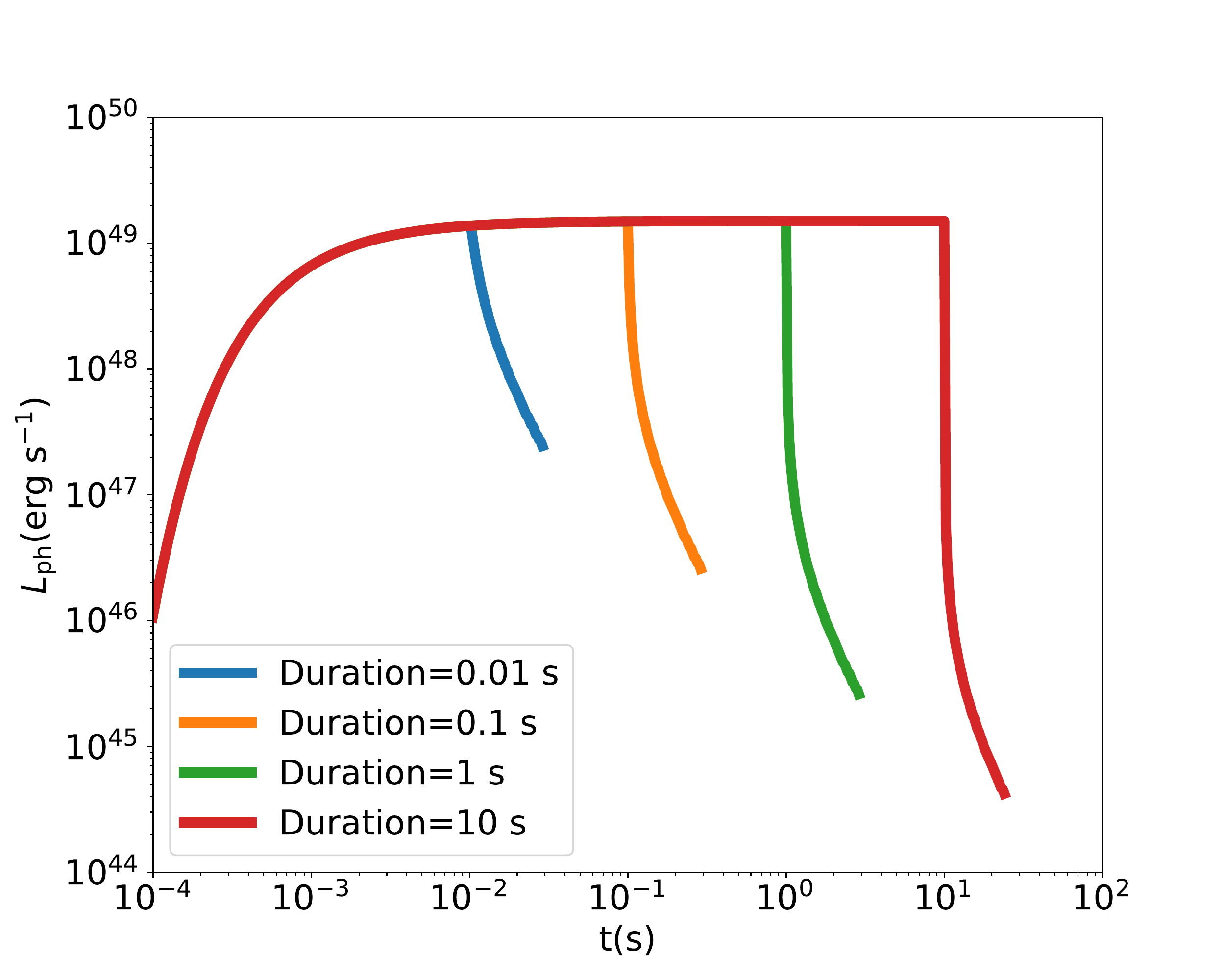}\put(-170,160){(b) the unsaturated emission}\\
   \includegraphics[width=\columnwidth]{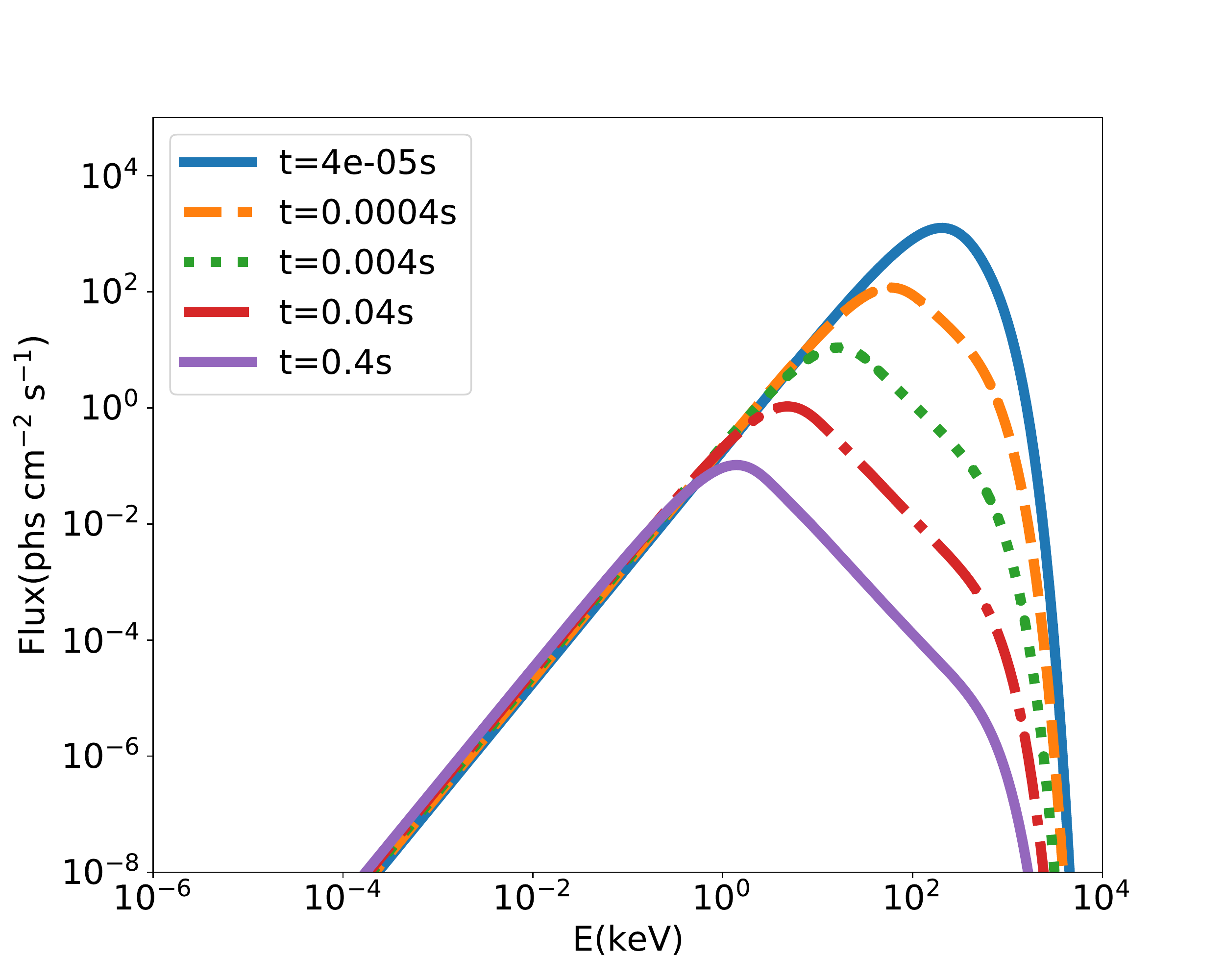}\put(-150,160){(c) the saturated emission}
  \includegraphics[width=\columnwidth]{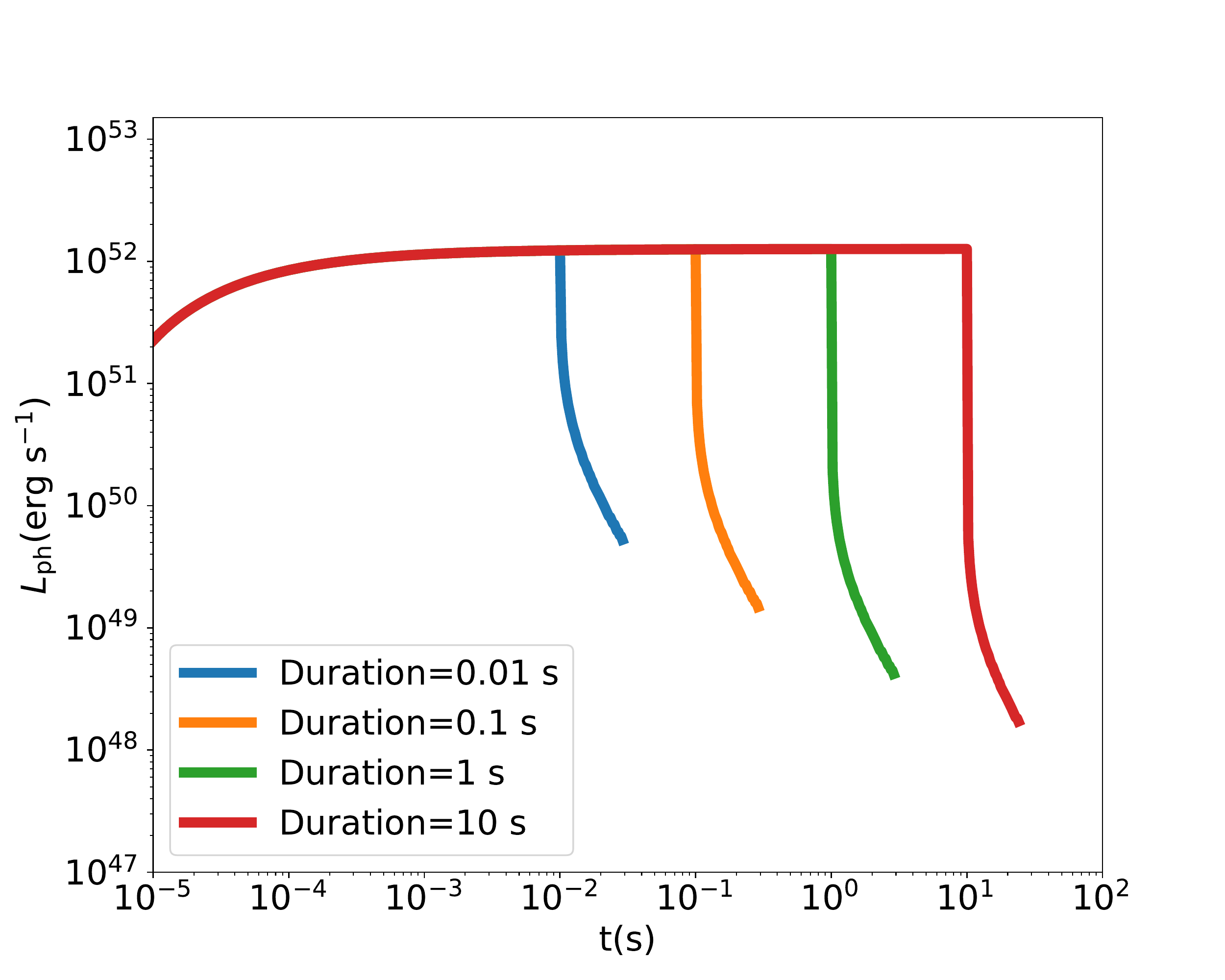}\put(-170,160){(d) the saturated emission}\\
\caption{(a) and (c) are the instantaneous photospheric spectra from the impulsive
injection for  the unsaturated emission with $L_{\rm w}=1.0L_{\rm w,49}$ and the saturated emission with $L_{\rm w}=1.0L_{\rm w,52}$ respectively, and the parameters are the same as those clarified in first paragraph Section~\ref{sec:caseII}. (b) and (c) are the photosphere luminosity light curves of continuous winds of the unsaturated and saturated emission, with $t_{\rm D}=$0.01s, 0.1s, 1s, 10s. The other parameters are the same as those clarified in first paragraph Section~\ref{sec:caseII}. \label{fig:Lph}}
\end{center}
\end{figure*}


\label{lastpage}
\end{document}